\documentclass[aps,pre,preprint,groupedaddress,showpacs,amsmath,amssymb,nofootinbib]{revtex4-1}
\usepackage{graphicx}% Include figure files
\usepackage{dcolumn}% Align table columns on decimal point
\usepackage{bm}% bold math
\usepackage{color}
\usepackage{epstopdf}

\newcommand{\pSG}{p_\text{SG}}

\begin{document}

\title{Entropic long range order in a 3D spin glass model}

\author{Maria Chiara Angelini$^1$ and Federico Ricci-Tersenghi$^2$}

\affiliation{$^1$ Dipartimento di Fisica, Universit\`{a} La Sapienza,
  P.le A. Moro 5, 00185 Roma, Italy\\
  $^2$ Dipartimento di Fisica, INFN -- Sezione di Roma 1 and CNR --
  IPCF, UOS di Roma,\\
  Universit\`{a} La Sapienza, P.le A. Moro 5, 00185 Roma, Italy}

\begin{abstract}
  We uncover a new kind of \emph{entropic} long range order in finite
  dimensional spin glasses.  We study the link-diluted version of the
  Edwards-Anderson spin glass model with bimodal couplings ($J=\pm 1$)
  on a 3D lattice.  By using \emph{exact} reduction algorithms, we
  prove that there exists a region of the phase diagram (at zero
  temperature and link density low enough), where spins are long range
  correlated, even if the ground states energy stiffness is null. In
  other words, in this region twisting the boundary conditions cost no
  energy, but spins are long range correlated by means of pure
  entropic effects.
\end{abstract}
  
% 75.50.Lk Magnetic properties of materials: Spin glasses and other
% random magnets

\maketitle

The low temperature phase of frustrated spin models is a very
interesting and debated subject \cite{DiepBook}.  Especially in models
with discrete couplings, lowering the temperature the frustration may
produce surprising effects.  For example, the classical ``order by
disorder'' effect discovered by Villain \textit{et al.} \cite{Villain}
shows up in 2D frustrated spin systems, where ground states (GS) have
no magnetization, while a spontaneous magnetization is present at any
positive temperature smaller than the critical one, $0\!<\!T\!<\!T_c$
(and this is a rather counterintuitive result!).  In this case, the
explanation is simple: due to the frustration, two subset of GS exist,
having ferromagnetic and antiferromagnetic long range order
respectively; these GS have exactly the same energy and so, at zero
temperature ($T=0$), they perfectly compensate each other, leading to
null global magnetization; nonetheless, at positive temperatures, the
energy of the ferromagnetic state is lower than the antiferromagnetic
one and a long range (ferromagnetic) order is recovered.  This example
evidences the importance of exact cancelations at $T=0$ in frustrated
models.

Among frustrated models, spin glasses (SGs) \cite{SGbooks} have a very
complex low temperature phase.  Entropy fluctuations in SGs with
discrete couplings are known to play an important role and are most
probably the main mechanism for making the free-energy spectrum
gapless \cite{KM_JLMM}.

In this work we study 3D spin glasses with binary couplings ($J=\pm
1$) at $T=0$, showing that frustration in SGs generates an effect even
more impressive than the one found by Villain \textit{et al.}: a long
range order only due to \emph{entropic} effects.  More precisely, in
this entropically ordered SG phase, a typical SG sample has many GS
with exactly the same energy, such that, summing over all these GS, no
long range order is found in the system at $T=0$. However, at a closer
look, all these GS are not really equivalent and taking into account
also the entropic contribution to the $T=0$ exact computation, we find
that a subset of GS is dominating the Gibbs measure and thus leads to
long range order in the system.

In order to explain the entropic long range order with simpler words,
we consider a pair of spins, $\sigma_i$ and $\sigma_j$, at a very
large distance, $|i-j| \simeq L$ (being $L$ the system size) and try
to estimate their thermodynamic correlation $\langle \sigma_i \sigma_j
\rangle$ at $T=0$ by computing the probabilities of being parallel or
antiparallel, $\mathbb{P}[\sigma_i = \pm \sigma_j]$.  The method,
which is typically employed, computes the GS energy at fixed
(relative) values of $\sigma_i$ and $\sigma_j$: if the resulting GS
energy difference $|E_\text{GS}(\sigma_i = \sigma_j) -
E_\text{GS}(\sigma_i = -\sigma_j)|$ (the so-called energy stiffness)
does not grow with $L$ the system is believed to have no long range
order.  But this conclusion is wrong!  Indeed, even if
$E_\text{GS}(\sigma_i = \sigma_j) = E_\text{GS}(\sigma_i =
-\sigma_j)$, the relative orientation of $\sigma_i$ and $\sigma_j$
still depends on the \emph{number} of GS, $\mathcal{N}_\text{GS}$,
with given values of $\sigma_i$ and $\sigma_j$:
\begin{equation*}
  \mathbb{P}[\sigma_i = \pm \sigma_j] \propto
  \mathcal{N}_\text{GS}(\sigma_i = \pm  \sigma_j) \propto
  \exp[S_\text{GS}(\sigma_i = \pm \sigma_j)]\;,
\end{equation*}
where $S_\text{GS}$ is the GS entropy.  If the entropy difference
$|S_\text{GS}(\sigma_i = \sigma_j) - S_\text{GS}(\sigma_i =
-\sigma_j)|$ grows with $L$, then $|\langle \sigma_i \sigma_j \rangle|
\to 1$ in the thermodynamical limit and the system shows an entropic
long range order (the energy stiffness being null).  Please note that
the present entropic effect is taking place also at $T=0$, while the
Villain's ``order by disorder'' requires a positive temperature
because it is due to an energy difference.

We are going to show, by exact reduction algorithms, that such an
entropic long range order exists in SGs with discrete couplings on
regular lattices in finite dimensions.  We consider a link-diluted 3D
Edwards-Anderson model defined by the Hamiltonian $H = -\sum_{<ij>}
\sigma_i J_{ij} \sigma_j$, where the sum is over all the nearest
neighbor pairs of a 3D simple cubic lattice of length $L$. The
couplings $J_{ij}$ are quenched, independent and identically
distributed random variables extracted from the distribution
\begin{equation}
P_J(J) = (1-p)\delta(J) +
\frac{p}{2}\big[\delta(J-1)+\delta(J+1)\big]\;,
\label{eq:distribuzione0}
\end{equation}
where $p \in [0,1]$ is the density of non-zero couplings.

This model has a critical line in the $(p,T)$ plane that separates the
paramagnetic phase from the SG phase. It was already shown by Bray and
Feng \cite{BrayFeng} that, while in a model with a continuous
couplings distribution this critical line ends for $T=0$ at the
geometric link percolation threshold $p_c$ \cite{pc}, for discrete
couplings the paramagnetic phase does extend beyond $p_c$, because of
exact cancellations between positive and negative couplings.  Let us
call $\pSG$ the critical value separating the paramagnetic from the SG
phase at $T=0$.  A tentative estimation of $\pSG$ has been provided by
Boettcher \cite{Boettcher} by considering the ``defect'' energy
$\Delta E_\text{GS}$ between the GS energies obtained by swapping
between periodic and anti-periodic boundary conditions along one
direction.  He found that for $p>p^* = 0.272(1)$ the variance of
$\Delta E_\text{GS}$ grows with $L$ (the mean being null by symmetry)
thus leading to a SG long range order. After the work of Boettcher the
threshold $p^*$ has been identified with $\pSG$, but this is not
generally true (as we are going to show now).  In general only the
inequality $\pSG \le p^*$ holds.  Recently in Ref.~\cite{JRT} the same
model has been solved exactly on the hierarchical lattice, showing
that $T=0$ computations can lead to misleading results.  Indeed, while
at $T=0$ the model shows a phase transition at $p^*$, the exact
solution at positive temperatures predicts a critical line in the
$(p,T)$ plane ending in $(\pSG,0)$, with $p_c<\pSG<p^*$ (strict
inequalities hold). The right critical point $\pSG$ is clearly
sensitive to entropic effects, that are neglected in the computation
of $p^*$.  The determination of the $\pSG$ value can be done by simply
considering first order corrections in temperature to the $T=0$
computations.  Thus, in the rest of the paper, we are going to work
in this $T=0^+$ limit.

On a 3D cubic lattice the model can not be solved exactly and Monte
Carlo methods are very inefficient at low temperatures.  To determine
the right critical point $\pSG$, we are going to apply some exact
decimation rules that reduce the system to a much smaller size, which
can be then easily solved by numerical methods.

We consider periodic boundary conditions in $x$ and $y$ directions,
while spins in $z=0$ and $z=L-1$ are linked respectively to two
different external spins, with quenched, independent and identically
distributed random couplings extracted from the distribution in
Eq.~(\ref{eq:distribuzione0}).  The addition of these external spins
does not modify the thermodynamic limit but it is very useful: to
check for percolation will be enough to find a path of non-zero
couplings between these two external spins, while to check for the
presence of long range order one can just measure the correlation
between these two spins.  So, in general, one will be satisfied with
the computation of the effective coupling between the two external
spins.

Given that the model is link diluted, we can eliminate recursively
weakly connected spins, generalizing what was done in
Refs.~\cite{Boettcher,JRT}.  In the original model couplings are
$T$-independent, but, by decimating spins, effective couplings are
created whose intensity will depend on temperature.  If we want to
find entropic effects, the first order correcting term in $T$ can not
be neglected, even studying the system in the $T=0$ limit.  For
infinitesimal $T$, we can write an effective coupling as
$J=\text{sign}(I)(|I|-TK)$ if $I\neq 0$ or $J=TK$ if $I=0$, where $I$
and $K$ are the energetic and entropic coupling respectively. The
choice for the relative sign is dictated by the fact that thermal
fluctuations decrease the coupling intensity.  Spins and bonds are
decimated using the following 5 rules.\\
{\bf R1} A zero- or one-connected spin is eliminated.\\
{\bf R2} A two-connected spin $\sigma$ is eliminated and an effective
coupling $J_{12}$ is created between the two neighboring spins,
$\sigma_1$ and $\sigma_2$, satisfying the equation
\[
\sum_{\sigma=\pm1} e^{(J_1\sigma\sigma_1+J_2\sigma\sigma_2)/T} \equiv
A e^{J_{12}\sigma_1\sigma_2/T}\;,
\]
for any choice of $\sigma_1$ and $\sigma_2$.  Expanding at first order
in $T$ the two members, we have for the energetic component
\[
I_{12} = \frac{1}{2}\Big(|I_1+I_2|-|I_1-I_2|\Big)\;,
\]
and for the entropic component
\begin{eqnarray*}
  K_{12} = K_1 & \quad\text{if}\quad & |I_1|<|I_2|\;,\\
  e^{2K_{12}} = e^{2K_1} + e^{2K_2} & \text{if} &
  |I_1|=|I_2| \neq 0\;,\\
  \tanh(K_{12}) = \tanh(K_1) \tanh(K_2) & \text{if} &
  I_1 = I_2 = 0\;.
\end{eqnarray*}
{\bf R3} Two bonds $J_{ij}^{1}$ and $J_{ij}^{2}$ between two spins $i$
and $j$ can be replaced by an effective coupling $J_{ij}$ with
components $I_{ij}=I_{ij}^{1}+I_{ij}^{2}$ and
$K_{ij}=K_{ij}^{1}+K_{ij}^{2}$.\\
{\bf R4} A three-connected spin $\sigma$ is eliminated and effective
couplings are created between the three neighboring spins
$\sigma_1$, $\sigma_2$ and $\sigma_3$, satisfying the equation
\[
\sum_{\sigma=\pm1} e^\frac{J_1 \sigma \sigma_1 + J_2 \sigma \sigma_2 +
  J_3 \sigma \sigma_3}{T} \equiv A e^\frac{J_{12}\sigma_1\sigma_2 +
  J_{23}\sigma_2\sigma_3 + J_{31}\sigma_3\sigma_1}{T}\;,
\]
for any choice of $\sigma_1$, $\sigma_2$ and $\sigma_3$.  Expanding at
first order in $T$ the two members, and introducing the couplings
$\widetilde{J}_0 = J_1 + J_2 + J_3$ and $\widetilde{J}_k =
\widetilde{J}_0 - 2 J_k$ with $k=1,2,3$, we get for the energetic
components
\begin{eqnarray}
  I_{12} & = & \frac{1}{4} \Big(|\widetilde{I}_0| - |\widetilde{I}_1|
  - |\widetilde{I}_2| + |\widetilde{I}_3|\Big)\;,\\
  I_{13} & = & \frac{1}{4} \Big(|\widetilde{I}_0| - |\widetilde{I}_1|
  + |\widetilde{I}_2| - |\widetilde{I}_3|\Big)\;,\\
  I_{23} & = & \frac{1}{4} \Big(|\widetilde{I}_0| + |\widetilde{I}_1|
  - |\widetilde{I}_2| - |\widetilde{I}_3|\Big)\;,
\end{eqnarray}
and for the entropic components
\begin{eqnarray}
  K_{12} = \frac{1}{4} \Big(f(\widetilde{J}_0) - f(\widetilde{J}_1)
  - f(\widetilde{J}_2) + f(\widetilde{J}_3)\Big)\;,\\
  K_{13} = \frac{1}{4} \Big(f(\widetilde{J}_0) - f(\widetilde{J}_1)
  + f(\widetilde{J}_2) + f(\widetilde{J}_3)\Big)\;,\\
  K_{23} = \frac{1}{4} \Big(f(\widetilde{J}_0) + f(\widetilde{J}_1)
  - f(\widetilde{J}_2) - f(\widetilde{J}_3)\Big)\;,
\end{eqnarray}
where $f(J)=|K|+\ln(1+e^{-2|K|})$ if $I=0$ and $f(J)=\text{sign}(I)K$
if $I \neq 0$.\\
{\bf R5} A spin $\sigma$ of any connectivity is eliminated if the
number $n_I$ of its couplings with a non-zero energetic component ($I
\neq 0$) does not exceed three ($n_I \le 3$). If $i,j=1,\ldots,n_I$
index the spins connected to $\sigma$ by couplings with $I \neq 0$,
and if $k$ indexes the other neighbours (for which $I_k=0$), then the
new couplings $J_{ij}$ are computed following previous rules, while
the new couplings $J_{ik}=\text{sign}(J_i)J_k$, i.e.\ $I_{ik}=0$ and
$K_{ik}=\text{sign}(I_i)K_k$.

We have applied recursively the above five rules in the order they are
listed: i.e., at each decimation step, we try to use rule R1, and,
only if it does not apply, we try to use rule R2, and if it does not
apply, we try to use rule R3, and so on.  The decimation process stops
when it reaches a reduced graph where none of the above five rules can
be applied.  This reduced graph does depend on the order the above
rules are applied (because rules R4 and R5 increase the degree of
neighboring spins), and the order we have chosen is the one producing
the smallest reduced graph.

If the couplings have a discrete spectrum then rule R3 may produce
exact cancellations, thus leading to null effective couplings: this is
the reason why $\pSG > p_c$ holds in general for models with discrete
couplings.  Applying the above rules recursively it is possible that,
starting with only energetic couplings, the final effective coupling
has only the entropic component (the energetic one being null).  In
this situation it is clear that entropic couplings are essential even
in the $T=0$ limit.

\begin{figure}[t]
\includegraphics[width=\columnwidth]{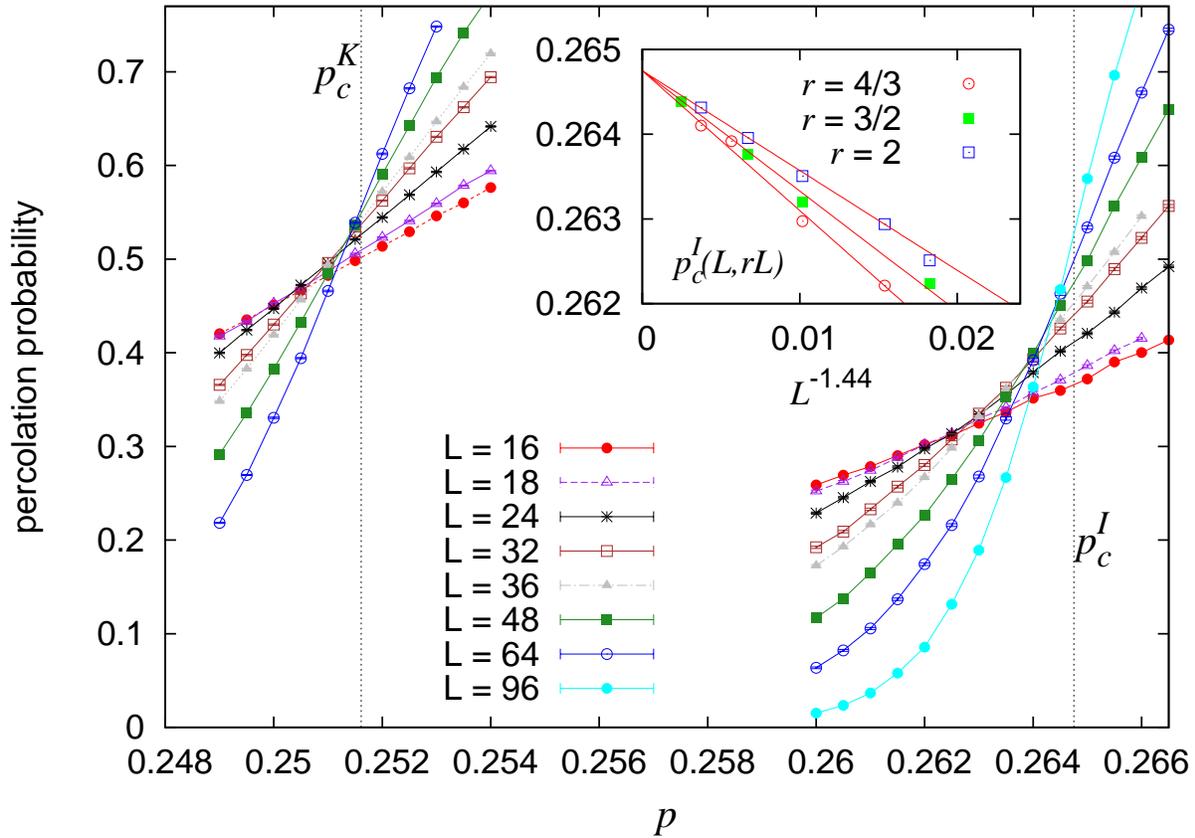}
\caption{\label{percolazioneinserto} Percolation probability for
  different lattice sizes $L$ as a function of the link density $p$
  for the energetic (right) and the entropic part (left) of the
  effective couplings. Inset: the infinite volume extrapolation for
  $p_c^I$.}
\end{figure}

First of all we study percolation properties of the networks of $I$
and $K$ effective couplings that result from the recursive application
of the above rules to all bond and spins (except the external ones,
that we want to keep).  We are mainly interested in the percolation
thresholds, $p_c^I$ and $p_c^K$, for the energetic and the entropic
components.  These percolation thresholds do depend on the set of
reduction rules and increase if more rules are used\footnote{The 5
  rules that we use are all those that keep the interactions
  pairwise. Indeed decimating a 4-connected spin would produce a
  4-spins effective interaction.}.  In Fig.~\ref{percolazioneinserto}
we show the percolation probabilities of the networks of $I$ and $K$
effective couplings for many different lattice sizes as a function of
the link density $p$.  By studying the crossing points $p_{L_1,L_2}$
of these probabilities for sizes $L_1$ and $L_2=r L_1$ with fixed $r$
(we use $r=3/4,\,3/2, \,2$) we have been able to estimate the
percolation thresholds $p_c^I$ and $p_c^K$ through fits including the
first scaling correction \cite{crossings}: $p_{L,r L} = p_c + A_r
L^{-1/\nu-\omega}$, as shown in the inset of
Fig.~\ref{percolazioneinserto}.  The resulting values are
$p_c^I=0.26475(10)$ and $p_c^K=0.25161(5)$.  The value of $p_c^I$ is
correctly lower than $p^*=0.272(1)$, the threshold value where a
positive energy stiffness emerges: the fact that $p_c^I$ and $p^*$ in
general differ can be easily understood by considering the 2D EA
model, which is clearly percolating, but has negative energy
stiffness.  Moreover $p_c^K$ is lower than $p_c^I$ because the applied
decimation rules leave the energetic component rational, while the
entropic one may become real, thus leading to much less exact
cancellations.  Please note that $p_c^K$ provides a lower bound to
$\pSG$ given that geometrical percolation of the effective couplings
is a necessary, but not sufficient, condition to have SG long range
order.

In the thermodynamic limit, for densities smaller than $p_c^I$, the
energetic component $I$ is not percolating and can not induce any long
range order.  Therefore, in the link density region $p_c^K < p <
p_c^I$ an eventual thermodynamic phase transition can be due solely to
entropic effects.

\begin{figure}[t]
\includegraphics[width=0.7\columnwidth]{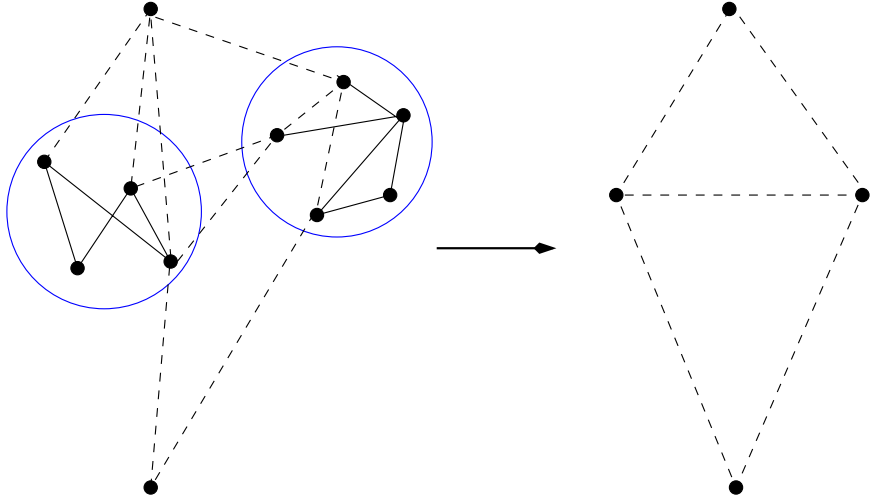}
\caption{\label{potts} On the left we show an example of the system
  after the decimation.  Full lines represent couplings with $I\neq0$,
  while dashed lines represent couplings with $I=0$.  The two
  $I$-clusters are enclosed in circles and are connected only by
  purely entropic couplings.  On the right, the system is mapped on a
  Potts model where each variable represents an $I$-cluster.  These
  Potts variables are connected by effective entropic couplings taking
  into account all the interactions originally connecting the
  $I$-clusters. Please note that our decimation rules always produce a
  reduced system with degrees not smaller than 4, but here we have
  drawn fewer lines for the sake of figure readability.}
\end{figure}

To search for such an entropic phase transition, we further reduce the
decimated graph.  For $p_c^K<p<p_c^I$, with high probability in the
large $L$ limit, the decimated graph is percolating solely in $K$,
while the $I$ couplings form clusters of finite size (similarly to
what happens in standard percolation below $p_c$). An example of the
resulting graph after the decimation process is shown on the left side
of Fig.~\ref{potts}, where full (resp. dashed) lines represent
couplings with (resp. without) a non-null energetic component $I$.
The two circles represent what we call $I$-clusters, that is groups of
spins connected by couplings with a non-null energetic component $I$
(note, however, that inside these $I$-clusters also couplings with
only the entropic component $K$ may exist, as in the rightmost circle
in Fig.~\ref{potts}). The connections between any two different
$I$-clusters have only entropic components.
 
Our idea is to map the original problem to a smaller and simpler one,
where the variables are the $I$-clusters, that interact only through
entropic couplings, as in the right side of Fig.~\ref{potts}

Given that we are interested in the $T=0$ limit, each $I$-cluster must
be in a ground state (GS) configuration.  So, for each $I$-cluster
$\mathcal{C}$, we compute with an exact branch\&bound algorithm all
its $\mathcal{N_C}$ GSs. We introduce then a Potts variable
$\tau_\mathcal{C}$ for that $I$-cluster, taking values in
$[1,\mathcal{N_C}]$.  We call
$\{\sigma_i^\mathcal{C}(\tau_\mathcal{C})\}$ the GS configurations of
the $\mathcal{C}$ cluster.

Working at $T=0$, the GSs are calculated by taking into account solely
the energetic component $I$ of the couplings. Afterward we consider
also the entropic components $K$, that give rise to two different
interacting terms.  $K$ bonds connecting two spins in the same
$I$-cluster produce a self-interaction term
\[
E^\mathcal{C}(\tau_\mathcal{C}) = \sum_{i,j \in \mathcal{C}} K_{ij}
\sigma_i^\mathcal{C}(\tau_\mathcal{C})
\sigma_j^\mathcal{C}(\tau_\mathcal{C})\;.
\]
This quantity may bias the choice among degenerate GSs even in the
$T=0$ limit. In the new Potts model, it can be interpreted like an
external field acting on the Potts variable $\tau_\mathcal{C}$ that
may bias its value.

$K$ bonds connecting spins in different $I$-clusters generate the
interaction between the Potts variables.  This interaction depends on
the configuration of both clusters, and so must be represented as a
matrix
\[
M^\mathcal{C,C'}(\tau_\mathcal{C},\tau_\mathcal{C'}) = \sum_{i \in
  \mathcal{C}, j \in \mathcal{C'}} K_{ij}
\sigma_i^\mathcal{C}(\tau_\mathcal{C})
\sigma_j^\mathcal{C'}(\tau_\mathcal{C'})\;.
\]
The Gibbs-Boltzmann measure for the reduced Potts model is then
\begin{equation}
  \mu\Big(\{\tau_\mathcal{C}\}\Big) \propto
  \exp\Big[ \sum_\mathcal{C,C'}
  M^\mathcal{C,C'}(\tau_\mathcal{C},\tau_\mathcal{C'}) +
  \sum_\mathcal{C} E^\mathcal{C}(\tau_\mathcal{C}) \Big]\;.
\label{Potts}
\end{equation}
It is important to note that this measure does not depend on the
temperature, because entropic couplings have a linear dependence on
$T$ that cancels the $1/T$ term in the Boltzmann factor.  The Potts
measure in Eq.~(\ref{Potts}) is an \emph{exact} effective description
of the original SG model at temperature $T=0^+$, having much less
variables and a smaller complexity with respect to the original model.

In order to locate a possible SG transition, we compute the
correlation between the external spins under the measure $\mu$ in
Eq.~(\ref{Potts}).  If the effective Potts model has a linear
topology, namely each variable has at most two neighbors, we solve it
exactly by the transfer matrix method (the probability $P$ to have
these ``linear'' systems is rather high: e.g., around the critical
density $\pSG$, $P>0.9$ for $L \le 24$, $P \simeq 0.7$ for $L=32$, $P
\simeq 0.6$ for $L=36$ and $P \simeq 0.2$ for $L=48$).  Otherwise we
use a Metropolis Monte Carlo method to sample the measure in
Eq.(\ref{Potts}), and the equilibration of the Markov chain is not an
issue given the small number of variables. Since the Gibbs-Boltzmann
measure in Eq.~(\ref{Potts}) does not depend on temperature, one can
think of it as that of a Potts model at $\beta=1$. Thus, for
equilibrating the corresponding Markov chain, we perform a simulated
annealing from $\beta=0$ to $\beta=1$, with steps $\Delta\beta=0.1$
and different cooling rates (100, 300 and 1000 Monte Carlo steps per
temperature).  We checked that the average of the interesting
quantities, like the correlations, does not dependend on the cooling
rate.

For the very few samples that show percolation in the energetic
components, we assume a correlation between external spins equal to
$1$.  This approximation makes no error in the thermodynamical limit
as long as $p<p_c^I$.

\begin{figure}[t]
\includegraphics[width=\columnwidth]{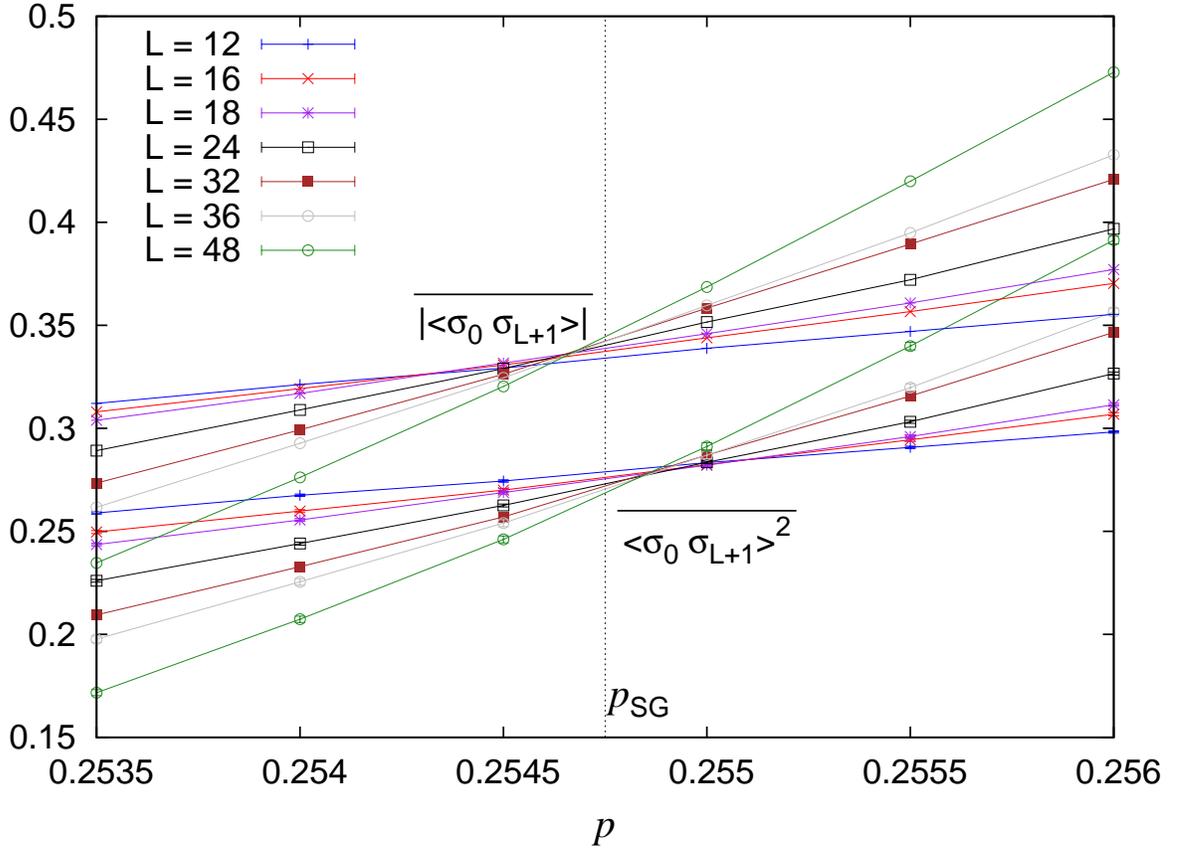}
\caption{\label{correlazione} Average of square (below) and absolute
  value (above) of correlations at distance $L$, for different lattice
  sizes, as a function of link density $p$. Errors are not larger than
  symbols.}
\end{figure}

\begin{figure}[t]
\includegraphics[width=\columnwidth]{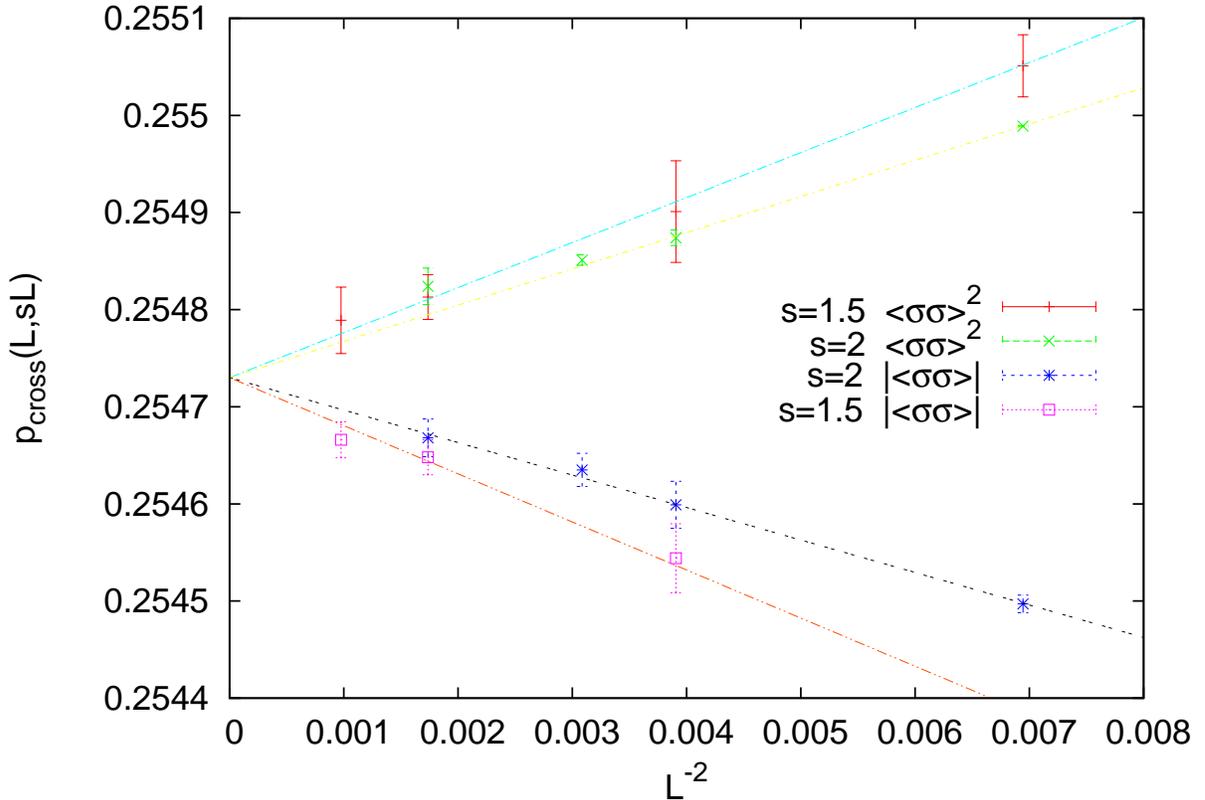}
\caption{\label{estrapolazione} Crossing points of data shown in
  Fig.~\ref{correlazione} with sizes $L$ and $sL$ as a function of
  $L^{-2}$.}
\end{figure}

Being interested in a SG long range order, we show in
Fig.~\ref{correlazione} the average over the samples of the square and
of the absolute value of the correlation between the external spins
(which are at distance $L$ in the original model) as a function of the
link density $p$. This quantity should decrease with $L$ in a
paramagnetic phase, while it should grow with $L$ if a SG long range
order is present: thus the crossing point of the curves in
Fig.~\ref{correlazione} roughly identifies the critical density
$\pSG$.  Our best estimation for $\pSG$ has been obtained by the
finite-size analysis of the crossing points of the correlations
measured in systems of sizes $L$ and $sL$, that should scale as
\[
p_\text{cross}(L, sL) = \pSG + B_s L^{-1/\nu-\omega}\;.
\]
In Fig.~\ref{estrapolazione} we show the values of $p_\text{cross}$
obtained with $s=1.5$ and $s=2$, together with the best fits.  In the
abscissa we have used the scaling variable $L^{-2}$ that provides the
best joint fit to all the data shown in the figure.  However the
uncertainty on this scaling exponent is large given the very small
spread of $p_\text{cross}$ around $\pSG$ for the sizes we have
studied.  Our final estimation for the SG critical threshold is $\pSG
= 0.25473(3)$.

By studying the slopes of the data shown in Fig.~\ref{correlazione} at
the critical point $\pSG$ as a function of the system size we have
been able to obtain an estimation of the exponent $\nu$ controlling
the shrinking of the critical region and we get $\nu=0.9(1)$.  This
value for the $\nu$ exponent does not coincide with the one measured
at criticality for the undiluted ($p=1$) or weakly diluted ($p=0.45$)
EA model, which is $\nu_T=2.2$ (the subscript $T$ should remind us
that this exponent is related to the shrinking of the critical region
in temperature).  However a simple argument gives the connection
between the two exponents: if the critical line close to the $T=0$
fixed point behaves like $T_\text{SG}(p) \propto (p-\pSG)^\phi$, then
$1/\nu_T=\phi/\nu$.  Our results thus suggest a value $\phi \simeq
0.4$ for the shape of the critical line.

\begin{figure}[t]
\includegraphics[width=0.6\columnwidth]{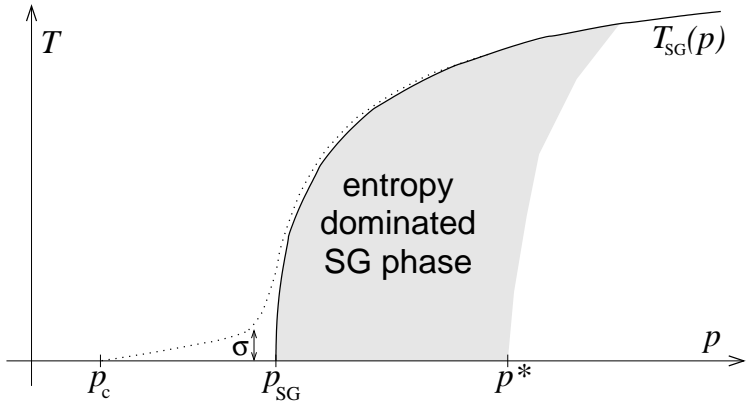}
\caption{\label{schematic} A schematic phase diagram in the $(p,T)$
  plane, showing that entropic long range order must exists also at
  positive temperatures. The dotted curve is the critical line of a
  model with quasi-discrete coupling.}
\end{figure}

We have shown that in 3D spin glasses frustration and coupling
discreteness may induce an entropic long range order: in this phase
the energy stiffness is zero (i.e.\ boundary conditions can be changed
at no energy cost), but the states with largest entropy dominate the
Gibbs measure. This dramatic effect of entropic contributions to the
Gibbs states has been extensively studied in mean-field models of spin
glasses with finite connectivity at $T=0$, especially in the contest
of random constraint satisfaction problems \cite{PNAS,coloring,KSAT}.
However in the present work we have proved the existence of such an
entropic phase in a 3D spin glass model.  Moreover this \emph{entropy
  dominated SG phase} should persist also at positive temperatures as
long as $p \lesssim p^*$ and the energy stiffness is null (see
Fig.~\ref{schematic}).

One may question that perfectly discrete couplings are difficult to
find in Nature.  Nonetheless if one considers a model with
quasi-discrete couplings (e.g. integer values plus a small Gaussian
term of variance $\sigma^2 \ll 1$) the critical line looks like the
dotted curve in Fig.~\ref{schematic}: it mainly follows the critical
line of the corresponding model with discrete couplings and only for
$T \lesssim \sigma$ moves towards $p_c$.  It is clear that the
identification of such a critical line is based on the correct
estimation of $\pSG$ in the model with discrete couplings.

\begin{figure}[t]
\includegraphics[width=0.6\columnwidth]{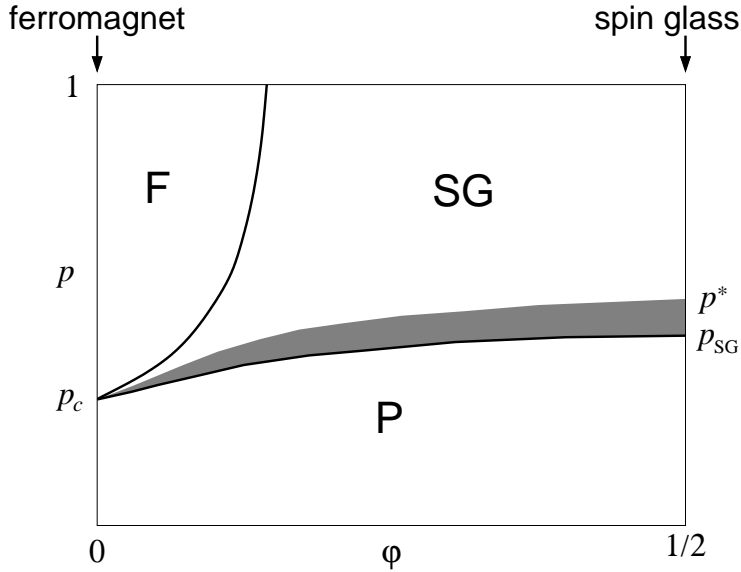}
\caption{\label{fig:ferroBias} Conjectured phase diagram at $T=0^+$
  by varying the level of frustration $\varphi$ in the model. The
  entropically long ranged phase should exist for any frustrated
  model, $\varphi>0$, with discrete energy levels.}
\end{figure}

One more comment about the generality of our results, regards what
happens when couplings have a ferromagnetic bias.  Indeed perfectly
symmetric coupling (i.e. with a null mean, $\overline{J_{ij}} = 0$)
are again difficult to find in Nature, and it is important to check
whether the entropic long range order is stable with respect to the
addition of a ferromagnetic bias in the couplings.  The answer is
contained in the pictorial phase diagram shown in
Fig.~\ref{fig:ferroBias}, where the link density $p$ is reported as a
function of some degree of frustration $\varphi$.  A quantitative
measure for $\varphi$ on a regular lattice can be, for example, the
fraction of frustrated elementary plaquettes: for $\varphi=0$ we have
a pure ferromagnetic model, while for $\varphi=1/2$ we have the spin
glass model studied in this work.  In this phase diagram, the addition
of a ferromagnetic bias in the couplings corresponds to reducing the
value of $\varphi$ with respect to the value $\varphi=1/2$ it takes in
a spin glass model with symmetrically distributed couplings.  The
phase diagram shown pictorially in Fig.~\ref{fig:ferroBias} contains,
in general, three different phases: a paramagnetic one (P), a spin
glass one (SG) and a ferromagnetic one (F).  Moreover, along the SG--F
boundary a mixed phase can exist \cite{CKRT}, containing a diverging
number of states with a non-null magnetization (but here we do not
want to enter the long-standing debate about the nature of the spin
glass phase in 3D models).  In Fig.~\ref{fig:ferroBias} the gray
region is our educated guess about the location of the entropically
long range ordered phase: in other words we conjecture the presence of
such a phase in any frustrated model ($\varphi > 0$) with discretized
energy levels.

An important comment regards the implications of our results on the
studies of the low temperature phase of SG models made by means of GS
computations. In these numerical studies one or few GS are usually
computed per sample, under different boundary conditions, and only the
GS energies are considered. Unfortunately this kind of study is not
able to identify the entropic long range order.  In light of our
results, this kind of numerical studies should be modified either
considering the first order correction in temperature when decimating
the variables, either computing many (or all) GS per sample, such as
to identify the state which is entropically dominating.  Some steps in
this direction have been already taken in \cite{HRT}, where it has
been recognized that a correct estimation of the GS clusters entropy
is necessary to extend predictions at positive temperatures.

Last, but not least, the present best estimation for the lower
critical dimension in SGs, $d_L \simeq 2.5$, is based on GS energy
stiffness computations \cite{Boettcher}, which ignore entropic
effects.  Most probably this result need to be modified to a lower
value due to the entropic long range order.

\end{document}